\documentclass[10pt,conference]{IEEEtran}
\IEEEoverridecommandlockouts 

\usepackage{cite}
\usepackage{amsmath,amssymb,amsfonts}
\usepackage{graphicx}
\usepackage{textcomp}
\usepackage{xcolor}
\usepackage{booktabs}
\usepackage{url}
\usepackage{hyperref}
\usepackage{tcolorbox}
\usepackage{listings}

\def\BibTeX{{\rm B\kern-.05em{\sc i\kern-.025em b}\kern-.08em
    T\kern-.1667em\lower.7ex\hbox{E}\kern-.125emX}}

\newcommand{\tool}{ImReq}

\usepackage{eso-pic}
\AddToShipoutPictureFG*{\AtPageLowerLeft{\put(\LenToUnit{0.08\paperwidth},\LenToUnit{0.022\paperheight}){\parbox{0.84\paperwidth}{\scriptsize\centering \textcopyright~2026 IEEE. Personal use of this material is permitted. Permission from IEEE must be obtained for all other uses, in any current or future media, including reprinting/republishing this material for advertising or promotional purposes, creating new collective works, for resale or redistribution to servers or lists, or reuse of any copyrighted component of this work in other works.}}}}

\begin{document}

\title{ImReq: An LLM-Powered Tool for Assessing and Improving Software Requirements Quality}

\author{\IEEEauthorblockN{Thanakorn Inthongsuk\IEEEauthorrefmark{1}, Pantaree Balee\IEEEauthorrefmark{1}, Jaosub Pongthawechai\IEEEauthorrefmark{1},\\ Chaiyong Ragkhitwetsagul\IEEEauthorrefmark{2}, Pattaraporn Sangaroonsilp\IEEEauthorrefmark{1}, Srisupa Palakvangsa-Na-Ayudhya\IEEEauthorrefmark{1}}
\IEEEauthorblockA{\IEEEauthorrefmark{1}Faculty of ICT, Mahidol University, Nakhon Pathom, Thailand\\
\IEEEauthorrefmark{2}Department of Computer Engineering, Faculty of Engineering, Chulalongkorn University, Thailand}}

\maketitle

\begin{abstract}
Requirements are a primary source of software needs and guide software teams in understanding what the system is expected to do. Ambiguous, incomplete, and unverifiable requirements remain a leading cause of software project failure. To promote high-quality requirements, ISO/IEC/IEEE 29148:2018 defines nine characteristics that a well-formed requirement should exhibit. Applying these characteristics manually, however, does not scale, while existing commercial AI-assisted requirements tools detect ambiguity without assessing conformance to the standard. In this paper, we present \tool{}, a web-based tool that evaluates natural-language requirements against all nine ISO/IEC/IEEE 29148:2018 characteristics and suggests improvements.
In a study with ten industry practitioners, \tool{} scored 73.50 on the System Usability Scale, and its improved requirements received a mean rank of 2.62/3.00 and were ranked best in 74\% of blind comparisons.

\smallskip
\noindent\textit{Replication package and demo video:} \url{https://github.com/MUICT-SERU/SP2026-ImReq}.
\end{abstract}


\section{Introduction}
\label{sec:intro}

Requirements are the contract between the customers and the development team, and defects introduced at this stage propagate into design, implementation, and test.
A survey of more than 2,000 practitioners by the Project Management Institute attributed 47\% of projects that missed their business goals to inadequate requirements management, and estimated a loss of USD~51 million for every USD~1 billion invested~\cite{pmi2014}.
A recent systematic review of 57 studies identified ambiguity as the leading cause of delayed analysis and project failure~\cite{khalid2025}.

The requirements engineering community mitigates these issues with standards and templates.
ISO/IEC/IEEE 29148:2018 (later referred to as ISO~29148 for brevity) enumerates nine characteristics of a well-formed individual requirement~\cite{iso29148} (listed in Table~\ref{tab:iso29148}), and the Easy Approach to Requirements Syntax (EARS) supplies five sentence templates that constrain how a requirement is phrased~\cite{ears2009}.
Femmer et al.~\cite{femmer2017} derived a catalog of \emph{requirements smells} from the natural-language criteria of ISO~29148 so that requirements could be checked as they are written rather than in a manual review.
Sahu et al.~\cite{sahu2024} went further and covered all nine characteristics with a rule-based Natural Language Processing (NLP) framework that condenses them into a single quality index. 
A systematic mapping of 105 empirical studies on requirements quality reports that existing research concentrates on ambiguity, completeness, consistency, and correctness; only nine studies validated their approaches in industrial settings with practitioners, and of the 41 tools developed, only 15 remained publicly available~\cite{montgomery2022}. Although ISO~29148 has been adopted to some extent in industry, its application remains inconsistent. A survey of 90 requirements engineers found that approximately 47\% were unfamiliar with ISO~29148, while those who were familiar with standards typically applied them based on personal decisions rather than organizational mandates~\cite{franch2022}. Recent studies have also demonstrated the potential of large language models (LLMs) to assess requirements quality and provide explanations for their judgments~\cite{lubos2024,krishna2024}. Lubos et al.~\cite{lubos2024} investigated the use of LLMs to evaluate requirements according to ISO~29148, while Krishna et al.~\cite{krishna2024} used LLMs to draft and review a software requirements specification (SRS). 

Beyond research tools, commercial requirements tools such as ScopeMaster~\cite{ScopeMaster}, Visure AI~\cite{VisureSolutions2026}, reqSuite~\cite{reqSuite}, and QVscribe~\cite{QRACorp} detect ambiguity and propose rewrites, and two recognize EARS patterns, but none assess requirements against the ISO~29148 characteristics.

In this paper, we present \tool{}, a web-based tool designed to address these gaps. \tool{} supports business and system analysts, requirements engineers, project managers, and developers in reviewing and improving software requirements.
Given a set of requirements, \tool{} evaluates each requirement against all nine ISO~29148 characteristics and provides a written rationale for each assessment. For requirements that do not conform to one or more characteristics, \tool{} generates improved versions following either ISO~29148 or EARS.

This paper makes two contributions:

\begin{itemize}
  \item an automated, publicly available tool that adopts all nine ISO~29148 characteristics to assess and improve natural-language requirements, with a workflow that keeps the adoption decision with the developers.
  \item an evaluation with ten industry practitioners covering usability, quality comparison, and suggestion quality.
\end{itemize}

\section{Background and Related Work}
\label{sec:background}

\begin{table}[t]
\caption{ISO~29148's characteristics of an individual requirement}
\label{tab:iso29148}
\centering
\footnotesize
\begin{tabular}{@{}lp{6.0cm}@{}}
\toprule
\textbf{Characteristic} & \textbf{The requirement\ldots} \\
\midrule
Appropriate  & is written at a level of detail suited to the
                            entity specified, and imposes no unnecessary
                            design constraints \\
Complete     & states all it needs to, requiring no further
                            amplification to be understood \\
Conforming   & follows the approved template and style
                            conventions of the requirement set \\
Correct      & accurately represents the stakeholder need
                            from which it was derived \\
Feasible     & can be realized within technical, cost, and
                            schedule constraints at acceptable risk \\
Necessary    & defines an essential capability, whose removal
                            would leave a deficiency \\
Singular     & states exactly one capability, with no
                            conjunctions bundling several together \\
Unambiguous  & admits only one interpretation \\
Verifiable   & can be confirmed by an accepted method:
                            inspection, analysis, demonstration, or test \\
\bottomrule
\end{tabular}
\end{table}

\textbf{Standards and templates.} ISO/IEC/IEEE 29148:2018 provides guidance for requirements engineering across the system life cycle. Section~5.2.5 in the standard defines nine characteristics of an individual requirement, as detailed in Table~\ref{tab:iso29148}. Six characteristics can be assessed from the requirement statement itself: Unambiguous, Singular, Complete, Verifiable, Appropriate, and Conforming. 
The remaining three characteristics require additional project-specific information: Correct, Feasible, and Necessary.

EARS complements the standard by prescribing how a requirement should be phrased, through five patterns: ubiquitous (\emph{the system shall}), event-driven (\textsc{when}), unwanted behavior (\textsc{if}\,/\,\textsc{then}), state-driven (\textsc{while}), and optional feature (\textsc{where})~\cite{ears2009}.
A previous study shows that applying EARS to industrial documents eliminated wordiness outright and reduced complexity and untestability~\cite{bigears2010}.

\textbf{Automated requirements quality assessment.} Moreno et al.~\cite{moreno2020} proposed an automated approach to assess and improve requirements quality by learning quality classifiers from expert-labeled requirements. Luitel et al.~\cite{luitel2023} used masked language models to identify omissions in natural-language requirements.
Sahu et al.~\cite{sahu2024} proposed a rule-based natural language processing framework for assessing the well-formed quality of individual requirements according to the nine characteristics defined in ISO~29148. The framework analyzes requirements at the token, span, syntactic, and sentence levels and combines the assessments of the nine characteristics into a well-formed requirement quality index.
A systematic review of 26 studies on AI-based requirements quality assessment found that most approaches score quality rather than improve it and rely on synthetic or public datasets~\cite{wolf2026}.
Krishna et al.~\cite{krishna2024} found GPT-4 capable of generating and validating SRS documents at a level comparable to entry-level engineers and providing constructive feedback for identifying and addressing requirements issues. The closest work to ours is by Lubos et al.~\cite{lubos2024}, who prompted Llama~2 (70B) once per requirement and per ISO~29148 characteristic, supplying the project scope and the characteristic's definition, and asked for a verdict, an explanation, and an improved requirement. The model found most of the issues the human raters did but over-flagged; its explanations were judged plausible in 70--100\% of cases, and its rewrites were judged improvements in most cases.
However, the study was performed manually, without any tool, and one dataset was artificially generated.

\section{The \tool{} Tool}
\label{sec:tool}

\begin{figure}[tb]
  \centering
  \includegraphics[width=0.8\columnwidth]{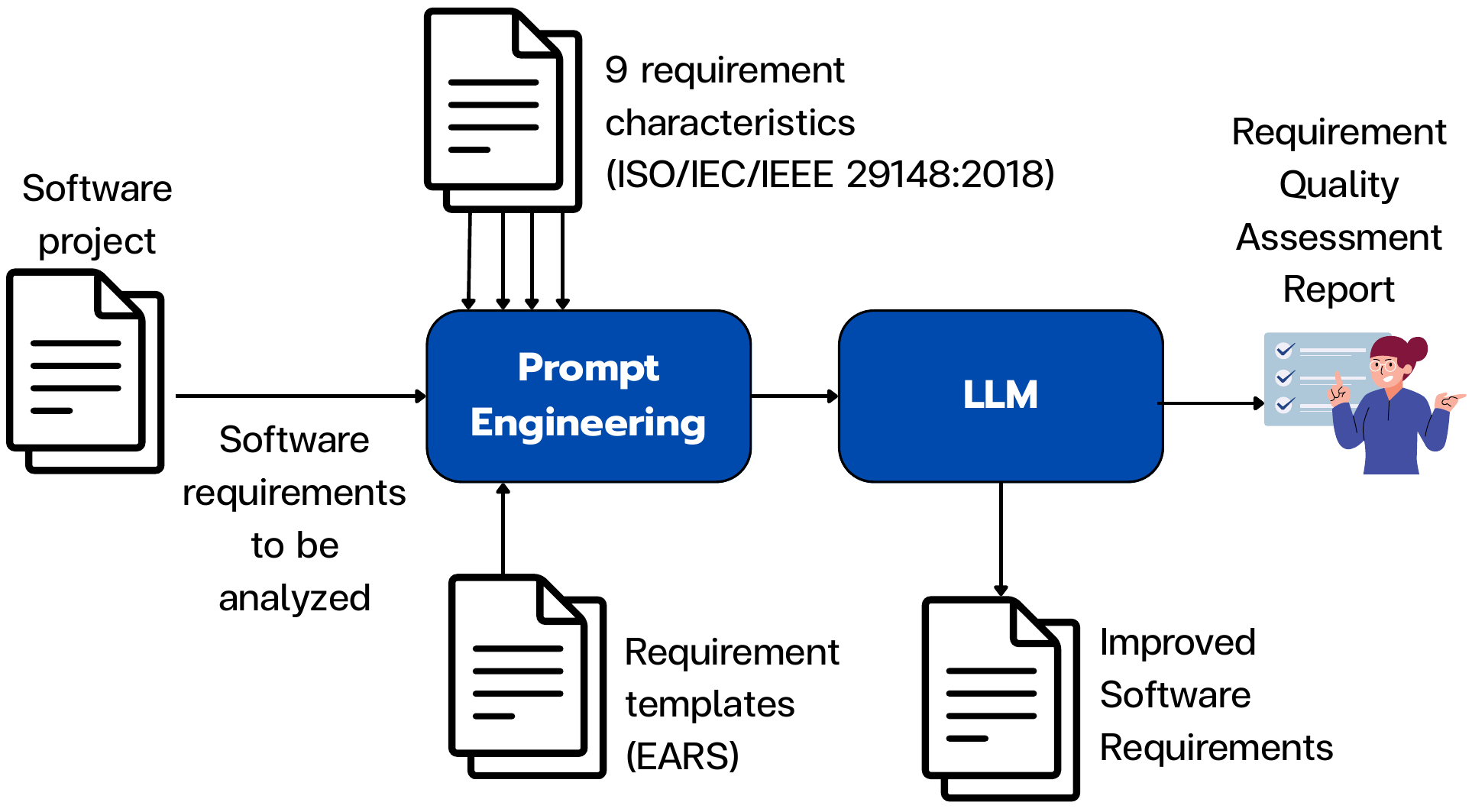}
  \caption{The key concept of ImReq}
  \label{fig:overview}
\end{figure}

The key concept of \tool{} is shown in Fig.~\ref{fig:overview}.
The tool accepts a set of requirements along with optional project reference documents. Then, it uses the nine ISO~29148 characteristics and the EARS templates, along with each requirement statement, to create prompts for the LLM. The LLM checks the requirement against each characteristic. If the requirement fails any characteristic, the LLM rewrites it using EARS or ISO~29148 sentence conventions. Finally, the tool produces the requirement quality assessment report with improved requirements.

\subsection{System Architecture} 
\label{sec:architecture}

\begin{figure}[tb]
  \centering
  \includegraphics[width=0.9\columnwidth]{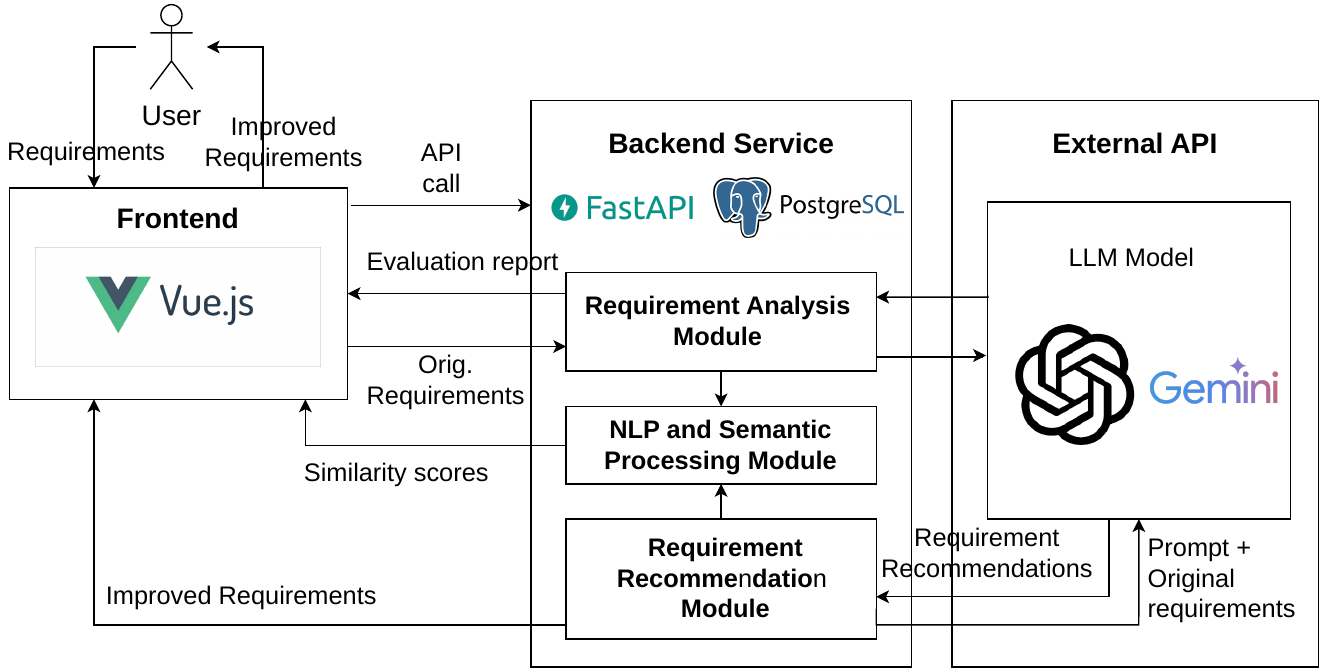}
  \caption{System architecture of \tool{}}
  \label{fig:architecture}
\end{figure}

The system architecture of \tool{} is depicted in Fig.~\ref{fig:architecture}.
The tool is a web application organized in three tiers.
A Vue.js single-page front end collects requirement sets and renders analysis results.
A FastAPI back end exposes the assessment and recommendation endpoints, and dispatches calls to an external LLM service asynchronously.
In the back end, a data layer stores users, projects, requirement versions, and analysis results in PostgreSQL, and holds a set of rule descriptions derived from ISO~29148 in a vector store that it queries when assembling a prompt.

The back end comprises three main modules.
\emph{Requirement Analysis} implements the per-criterion assessment.
\emph{Requirement Recommendation} produces improved requirements.
\emph{NLP and Semantic Processing} computes similarity indicators between original and rewritten requirements.
Requirement sets enter and leave \tool{} as CSV files.

\subsection{Per-Criterion Quality Assessment}
\label{sec:assessment}
\tool{} issues one request per characteristic and dispatches the nine requests concurrently.
Each request carries only the rule text and instructions relevant to a single characteristic.
This bounds the length of every prompt and confines each response to one criterion.

Each prompt is assembled from six parts.
A \emph{role} instruction casts the model as an ISO~29148 requirements-quality expert.
\emph{Rule injection} inserts the rule description for the characteristic under assessment.
\emph{Project-context injection} adds excerpts from the reference documents uploaded with the project, but only for the three characteristics that cannot be judged from the requirement sentence alone, i.e., Correct, Feasible, and Necessary.
An explicit \emph{decision rule} fixes the PASS/FAIL condition for that characteristic.
A \emph{JSON schema} constrains the reply so the back end can consume it without post-hoc parsing.
Finally, \emph{constraints} forbid the model from introducing numeric values or conditions absent from the source requirement, and require the reply to be written in the same language as the requirement.

Each request returns a verdict and, when the verdict is FAIL, a written rationale explaining the failure.
If the project reference documents are not given, \tool{} displays Correct, Feasible, and Necessary as UNDETERMINED.

\subsection{Requirement Recommendation}
\label{sec:recommendation}

For requirements that fail one or more characteristics, \tool{} generates an improved version based on two possible options: EARS or the ISO~29148 sentence conventions.
Rewriting addresses the specific deficiencies the assessment identified. For example, the rewriting supplies a missing subject and the modal verb \emph{shall}, replaces a capability fragment with a well-formed requirement sentence, adds an observable acceptance condition where verifiability was lacking, or decomposes a compound requirement into a set of singular sub-requirements.

\subsection{Meaning-Preservation Check}
\label{sec:similarity}

Rewriting requirements with a generative model may result in improved text that is clearer yet no longer keeps the original meaning.
\tool{} addresses this by reporting two complementary similarity measures between each original requirement and its replacement.
\emph{Token similarity} captures lexical overlap using the Jaccard index of the token sets, and
\emph{Meaning similarity} captures semantic correspondence, using SBERT sentence embeddings.
The similarity scores serve as supporting information. High similarity cannot rule out changes such as an inserted negation or an altered numeric value, so the final decision rests with the user.

\section{Usage Walkthrough}
\label{sec:usage}

\begin{figure}[t]
  \centering
  \includegraphics[width=0.95\columnwidth]{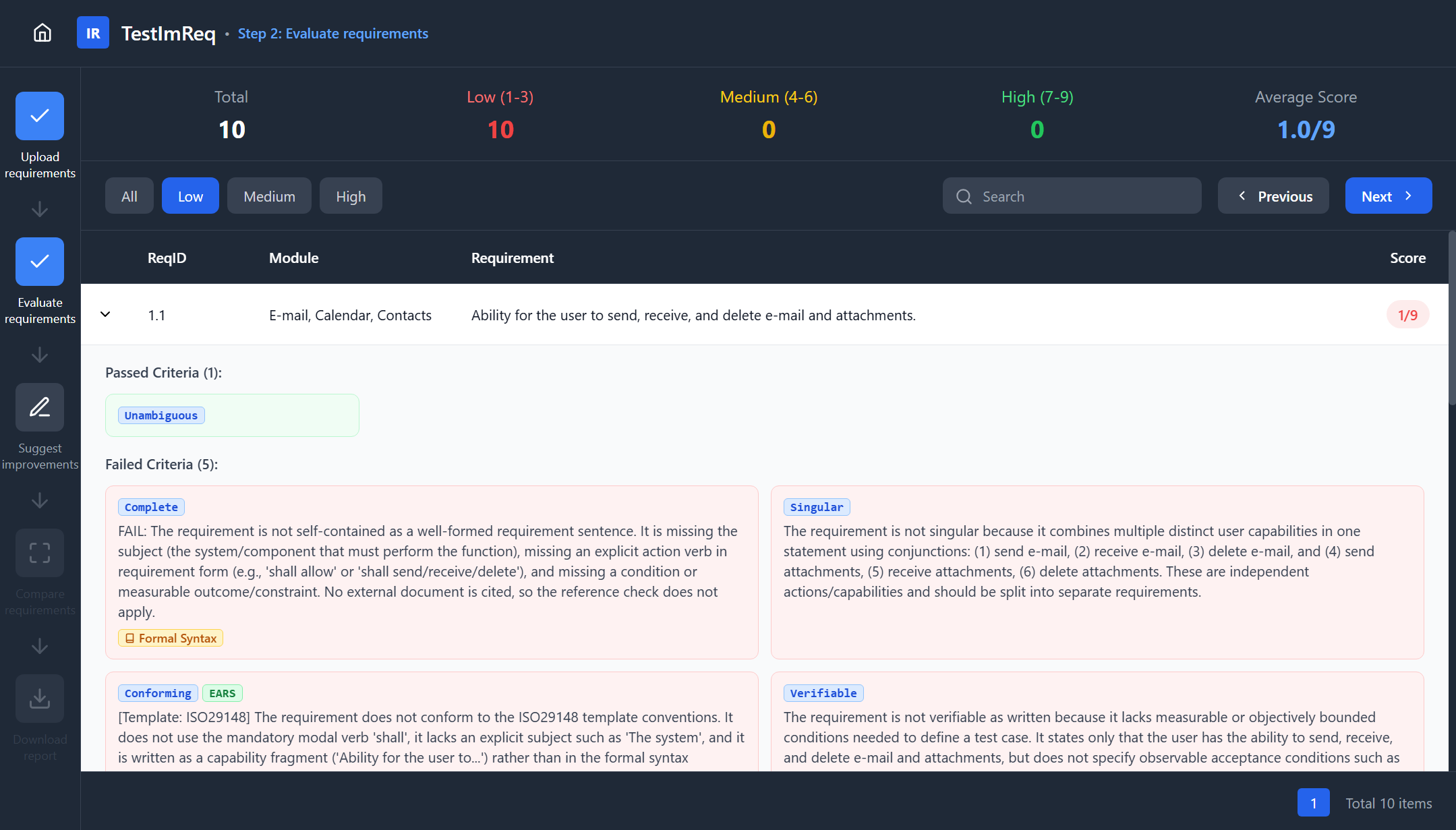}
  \caption{ImReq's requirement quality assessment report}
  \label{fig:evaluate}
\end{figure}

\begin{figure}[t]
  \centering
  \includegraphics[width=0.95\columnwidth]{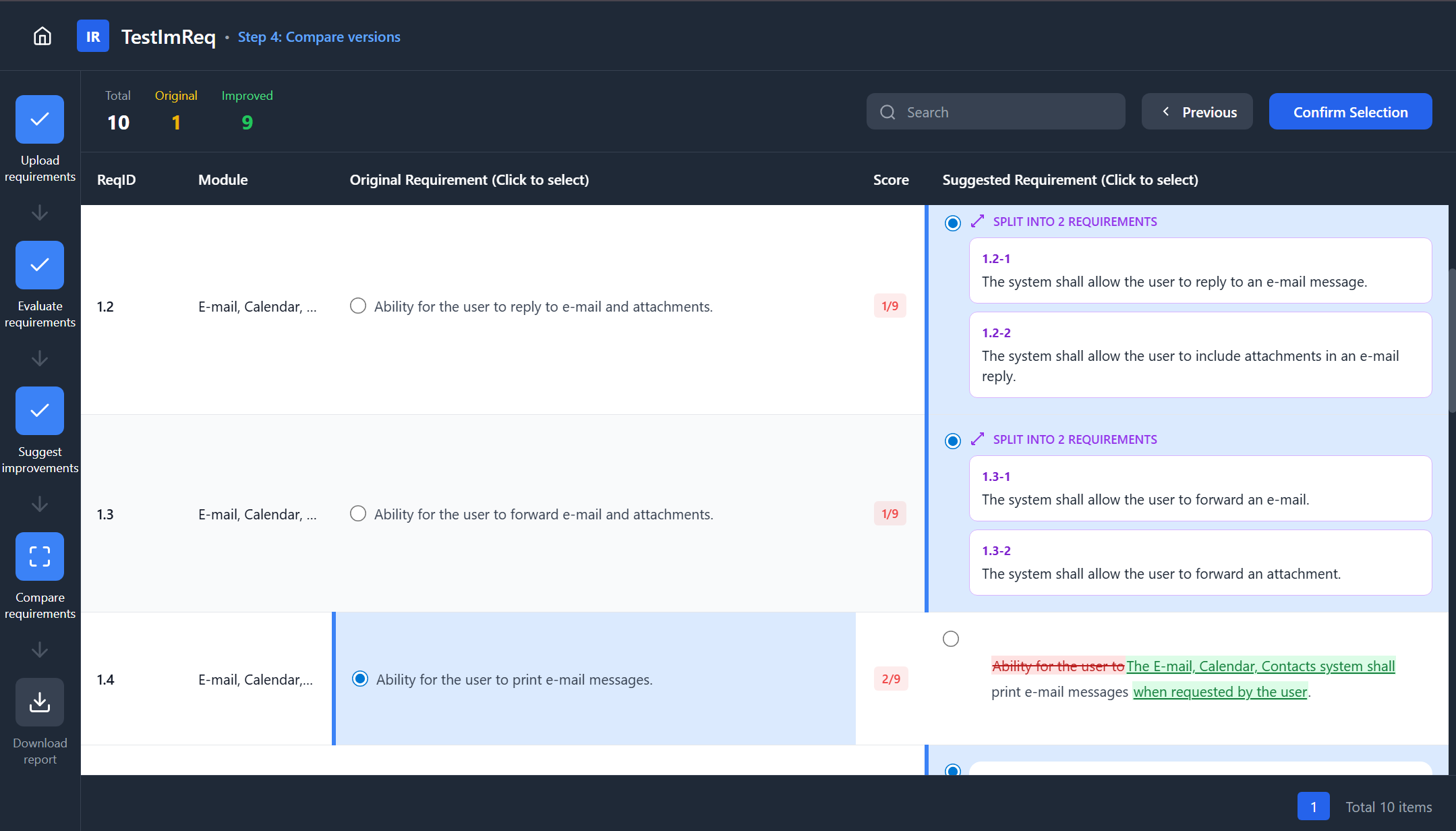}
  \caption{Original vs.~suggested versions of the requirements} 
  \label{fig:compare}
\end{figure}
We illustrate the workflow of \tool{} as follows.
First (project setup), the user creates a project, optionally uploads reference documents such as a project brief or domain glossary, and chooses the target template, EARS or ISO~29148.
Second, the user uploads a CSV file and maps its columns onto \tool{}'s fields: \textsf{Requirement} is required, while \textsf{ReqID} and \textsf{Module} are optional and generated if absent.
Third, \tool{} assesses every requirement against the nine characteristics.
A dashboard reports the Low/Medium/High distribution and the mean score, and each row expands to show passed and failed criteria with their rationales (Fig.~\ref{fig:evaluate}).
Our example requirement, \emph{``Ability for the user to send, receive, and delete e-mail and attachments''}, scores 1/9.
It passes Unambiguous, but fails Complete (no subject, no action verb in requirement form), Singular (six independent capabilities in one sentence), Conforming (a capability fragment rather than a \emph{shall} statement), and Verifiable (no observable acceptance condition).

Next, \tool{} rewrites the failing requirements, and the similarity summary flags rewrites whose wording or meaning drifts from the original for the user's review. Then, the original and suggested versions are shown side by side with the differences highlighted (Fig.~\ref{fig:compare}), and the user selects which version to keep for each requirement.
Lastly, the selected requirements are exported as CSV for use in the team's existing process.

\section{Evaluation}
\label{sec:evaluation}

\subsection{Selecting the LLM}

We compared ChatGPT (\texttt{gpt-5.4}) and Gemini (\texttt{gemini-3.5-flash}) as the engine behind \tool{}.
Fifty requirements drawn from five projects of the PURE dataset~\cite{Ferrari2017} were improved by both models, and three authors independently judged the resulting pairs blind to their source.
ChatGPT was preferred in 76 of 150 judgments (Gemini: 74) and won 28 of 50 per-requirement majority votes.
We thus made ChatGPT the default. \tool{} can use any LLM.

\subsection{User Study with Industry Practitioners}
We evaluated \tool{} with ten industry practitioners, mainly from SMEs in Thailand.
The participants included four software engineers, three project managers, two system analysts, and one CTO. Eight participants had more than five years of software development experience.
Each session lasted about 60 minutes and comprised informed consent, a pre-test questionnaire and a blind quality comparison, hands-on use of \tool{} with 10--15 of the participants' own requirements, and a closing System Usability Scale (SUS)~\cite{brooke1996} questionnaire and semi-structured interview.\footnote{The study was approved by the Mahidol University Central Institutional Review Board (MUCIRB 2026/074.1702).}
Based on the pre-test questionnaire, we found that every participant reported problems with ambiguous requirements and nine of ten with incomplete ones. Five spent more than five hours per project reviewing requirements.
None used a formal checklist, standard, or tool for reviews.

\subsection{Results}

\tool{} scored 73.50 on the SUS, which corresponds to \emph{Good}, higher than the conventional acceptability threshold of 68~\cite{bangor2009}.
Participants agreed most strongly that the tool was quick to learn (4.3/5.0) and reported high confidence in using it (4.1/5.0).
In the blind comparison, every participant received the same 10 requirements from the PURE dataset~\cite{Ferrari2017}. For each requirement, they saw three versions in random order with their sources hidden: (1) the original, (2) \tool{} with ChatGPT, and (3) \tool{} with Gemini. Participants ranked the three versions from 1 (worst) to 3 (best). Across the 100 rankings (10 participants $\times$ 10 requirements), \tool{} with ChatGPT was ranked best 74 times (74\%), \tool{} with Gemini 25 times, and the original once. The mean rank was 2.62 for \tool{} with ChatGPT.

Rating the suggestions directly, participants perceived that \tool{} correctly identified defects in their original requirements (4.1/5). Participants also judged that the rewrites preserved the original meaning (4.2/5), and that the tool reduced the time needed to review and revise requirements (4.2/5).
Grammatical clarity scored 3.8/5 and real-project usability 3.7/5.

\section{Threats to Validity}
\label{sec:threats}

\textbf{Construct.} SUS captures perceived usability after a single session, and the quality ratings may contain human biases. 
\textbf{Internal.} The LLM is non-deterministic, and we report a single pass, so run-to-run variance is unquantified.
\textbf{External.} The projects and participants are drawn from a single country with a small sample size, so generalization is limited.

\section{Conclusion}
\label{sec:conclusion}

\tool{} makes the nine ISO~29148 quality characteristics operational for practitioners, assessing each requirement per characteristic with an explanation, rewriting the failures into EARS or ISO~29148 form, and leaving the adoption decision with the user.
Ten practitioners rated it 73.50 on the SUS and ranked its improved requirements best among three blind versions in 74\% of comparisons.
Future work will address the over-splitting behavior of Singular characteristic, ground rewrites in a project glossary using established terminology, and evaluate the tool across multiple organizations.

\bibliographystyle{IEEEtran}
\bibliography{references}

\end{document}